# Front End for a Neutrino Factory or Muon Collider

**David Neuffer,**[a,*] **Pavel Snopok,**[b] **and Yuri Alexahin**[a]

[a] *Fermilab, PO Box 500, Batavia, IL 60510, USA*
[b] *Illinois Institute of Technology, Chicago, IL 60616, USA*
[*] *E-mail*: neuffer@fnal.gov

ABSTRACT: A neutrino factory or muon collider requires the capture and cooling of a large number of muons. Scenarios for capture, bunching, phase-energy rotation and initial cooling of μ's produced from a proton source target have been developed, initially for neutrino factory scenarios. They require a drift section from the target, a bunching section and a ϕ-δE rotation section leading into the cooling channel. Important concerns are rf limitations within the focusing magnetic fields and large losses in the transport. The currently preferred cooling channel design is an "HFOFO Snake" configuration that cools both $\mu^+$ and $\mu^-$ transversely and longitudinally. The status of the design is presented and variations are discussed.

# Contents

# 1. Introduction

Scenarios have been developed for using muons in a storage ring based neutrino source or "neutrino factory" and in a high-energy high-luminosity "muon collider" [1, 2, 3]. The scenarios are outlined in figure 1. In both scenarios high intensity proton bunches from a proton source strike a production target producing secondary particles (mostly $\pi^{\pm}$'s). The π's decay to μ's and the μ's from the production are captured, bunched, cooled and accelerated into a storage ring for neutrinos or high energy collisions. The present paper discusses the section of the scenarios labelled the "front end" in figure 1, between the target and the accelerator for the neutrino factory and between the target and "6-D" cooling section of the muon collider.

    In the Front End, pions from the production target decay into muons and are focused by magnetic fields and bunched by time-varying electric fields into an initial cooling system that forms the muons into a beam suitable for the following acceleration and/or cooling of the scenarios.

    μ's from the target and decay are produced within a very broad energy spread and length. Initially, capture within a single bunch was considered, but that requires either very low frequency rf (<20MHz) or novel induction linacs. The scale and cost of such a system would be uncomfortably large. Instead a novel system of higher frequency rf cavities (~200—500 MHz) was developed that forms the μ's into a train of manageable bunches, using current-technology rf cavities and power sources [4, 5]. The same system can be used for both neutrino factory and collider scenarios.

    In this paper we present this Front End system, with a detailed description of its most recent implementations within the MAP feasibility study. Features and variations incorporating

recent research are discussed and performance for neutrino factory / collider applications is described. These features include use of a chicane and absorber to reduce downstream particle losses and use of gas-filled rf to suppress breakdown.

The rf bunching naturally forms the beam for an initial cooling section. For the neutrino factory international design study (IDS) [6], this uses a simple solenoidal focusing system with LiH absorbers that provides only transverse cooling (4-D phase-space cooling). More recently a 6-D initial cooling system (the "HFOFO Snake") using tilted solenoids and LiH wedge absorbers was developed, and is currently considered somewhat superior, particularly for a collider scenario [7, 8]. The initial cooling concepts are discussed and compared.

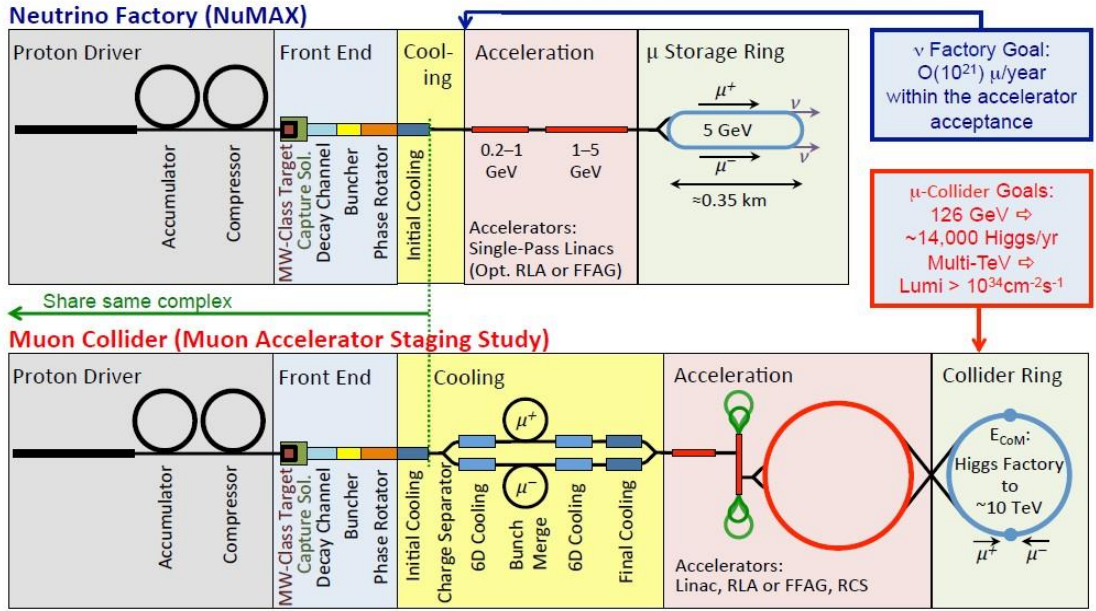

**FIGURE 1:** Block diagrams of neutrino factory and muon collider scenarios. The present paper discusses the "Front End" section of the scenarios from the Target to the Cooling/Acceleration sections. Discussion of the "Initial Cooling" section is also considered as essential in matching to the downstream facilities. In principle the same Front End can be used in both scenarios.

## 2. Front End overview – drift, bunching and phase rotation

The Front End concept presented here was generated for the Neutrino Factory design studies [9], and subsequently extended and reoptimized for the Muon Accelerator Program (MAP) muon collider design studies.[10, 11] The Front End system takes the π's produced at the target, and captures and initiates cooling of the resulting decay μ's, preparing them for the μ accelerators. Figure 2 shows an overview of the system, as recently developed for the MAP studies. In this figure, the transport past the target is separated into drift, buncher, rotator and cooling regions. Figure 3 shows an idealized view of the longitudinal phase space through the Front end region.

The reference input beam was a monochromatic 8 GeV proton beam with a 3 ns time spread incident on a liquid mercury target in the bore of a 20 T solenoid. The interaction of the beam with the target was modeled using the MARS15 (2010) default generator [12, 13].

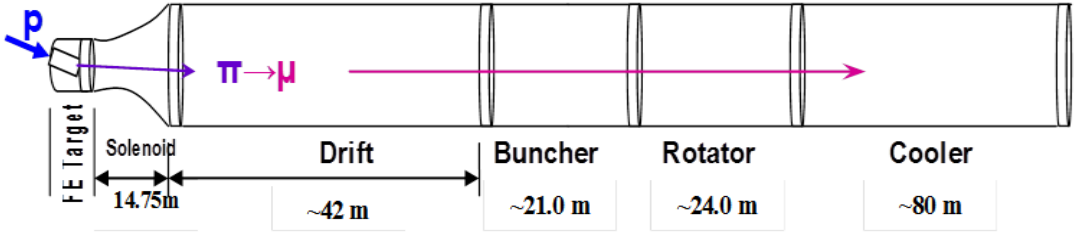

**Figure 2:** Overview of the MAP front end, consisting of a target solenoid (20 T), a capture solenoid (20 T to 2.0T, 14.75 m), Drift section (42 m), rf Buncher (21 m), an energy-phase Rotator (24 m), with a Cooler (80 m).

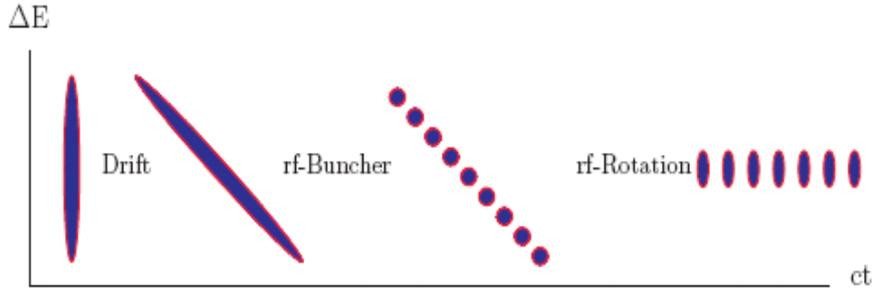

**Figure 3:** Overview of longitudinal motion in the Front End. An initial muon distribution with large energy spread and small bunch length stretches to a distribution with an energy-position correlation in the Drift. The rf-Buncher forms the beam into a string of different energy bunches and the rf-Rotator moves the bunches to equal energies, forming a string of bunches for the downstream Cooler.

### 2.1 Drift for decay

The multi-GeV proton source produces short pulses of protons that are focused onto a target immersed in a high-field solenoid with an internal beam pipe radius $r_{sol}$. The proton bunch length is 3 ns rms (~15 ns full-width), $B_{sol}$ =20 T, and $r_{sol}$ = 0.075 m, at initial baseline parameters. Secondary particles are radially captured if they have a transverse momentum $p_T$ less than ~$ecB_{sol}r_{sol}/2$ = 0.225 GeV/c. Downstream of the target solenoid the magnetic field is adiabatically reduced from 20T to 2T over ~14.75 m, while the beam pipe radius increases to ~0.25 m. This arrangement captures a secondary pion beam with a broad energy spread (~50 MeV to 400 MeV kinetic energy).

The initial proton bunch is relatively short, and as the secondary pions drift from the target they spread apart longitudinally, following: $c\tau(s) = s/\beta_z + c\tau_0$, where s is distance along the transport and $\beta_z = v_z/c$. Hence, downstream of the target, the pions and their daughter muons develop a position-energy correlation in the RF-free drift. In the MAP baseline, the total drift length is $L_D$ = 64.6 m, and at the end of the decay channel there are about 0.2 muons of each sign per incident 8 GeV proton.

### 2.2 RF Buncher

The drift channel is followed by a buncher section that uses rf cavities to form the muon beam into a train of bunches, and a phase-energy rotating section that decelerates the leading high-energy bunches and accelerates the later low-energy bunches to the same mean energy [5]. To determine the buncher parameters, we consider reference particles (0, N) at $P_0$= 250 MeV/c and $P_N = 154$ MeV/c, with the intent of capturing muons from a large initial energy range (~50 to ~400 MeV). The rf frequency $f_{rf}$ and phase are set to place these particles at the center of bunches while the rf voltage increases along the transport. This requires that the rf wavelength $\lambda_{rf}$ increases, following:

$$N_B \lambda_{rf}(s) = N_B \frac{c}{f_{rf}(s)} = s\left(\frac{1}{\beta_N} - \frac{1}{\beta_0}\right)$$

where $s$ is the total distance from the production target, $\beta_1$ and $\beta_2$ are the velocities of the reference particles, and $N$ is an integer. For the baseline, $N$ is chosen to be 12, and the buncher length is 21m. Therefore, the rf cavities decrease in frequency from ~490 MHz ($\lambda_{rf} = 0.61$ m) to ~365 MHz ($\lambda_{rf} = 0.82$m) over the buncher length.

The initial geometry for rf cavity placement uses 2 0.25 m long cavities placed within 0.75 m long cells. The 2T solenoid field focusing of the decay region is continued through the Buncher and the Rotator. The rf gradient is increased along the Buncher, and the beam is captured into a string of bunches, each of them centered about a test particle position with energies determined by the $\delta(1/\beta)$ spacing from the initial test particle:

$$1/\beta_n = 1/\beta_0 + n\, \delta(1/\beta),$$

where $\delta(1/\beta) = (1/\beta_N - 1/\beta_0)/N$. In the initial design, the cavity gradients follow a linear increase along the buncher:

$$V'_{rf}(z) \approx 15\left(\frac{z}{L_{Bf}}\right) MV/m$$

where $z$ is distance along the buncher and $L_{Bf}$ is the buncher section length. The gradual increase in voltage gradient enables a somewhat adiabatic capture of muons into separated bunches.

In a practical system, the varying frequency rf system would be implemented by a limited number of discrete frequency cavities. In our initial design 54 rf cavities using 14 different rf frequencies within the 490—366MHz range are used within the 21m long buncher. The rf gradients of the cavities increase from 0 to 15 MV/m over that length while the rf frequency decreases. Table 1 summarizes the rf cavity requirements. At the end of the buncher, the beam is formed into a train of positive and negative bunches of different energies.

TABLE 1. Summary of front end RF requirements

| Region | Length (m.) | Number of Cavities | Number of frequencies | Frequencies [MHz] | Peak Gradient [MV/m] | Peak Power requirements |
|---|---|---|---|---|---|---|
| Buncher | 21 | 54 | 14 | 490 to 366 | 0 to 15 | 1 to 1.34 MW/cavity |
| Rotator | 24 | 64 | 16 | 366 to 326 | 20 | 2.4 MW/cavity |
| Cooler | 75 | 200 | 1 | 325 | 25 | 3.7 MW/cavity |
| Total (with Drift) | 230m | 318 | 31 | 490 to 325 | 1700 MV | 1000 MW |

## 2.3 Rotator

In the rotator section, the rf bunch spacing between the reference particles is shifted away from the integer $N_B$ by an increment $\delta N_B$, and phased so that the high-energy reference particle is stationary and the low-energy one is uniformly accelerated to arrive at the high-energy at the end of the Rotator. For the IDS, $\delta N_B = 0.05$ and the bunch spacing between the reference particles is $N_B + \delta N_B = 12.05$. The Rotator consists of 0.75 m long cells with 2 0.25 m rf cavities at 20 MV/m. The rf frequency of cavities decreases from 365 MHz to 326 MHz down the length of the 42 m long rotator region. The rotator uses 64 rf cavities of 16 different frequencies (see table 1). At the end of the rotator the rf frequency matches into the rf of the ionization cooling channel (325 MHz).

For transverse focusing, the Solenoidal field of 2 T is maintained throughout the rotator and is matched into an alternating Solenoidal field for the cooler. The match is obtained by perturbations on the focusing coil currents of the first four cells of the cooler lattice. (Other field configurations that confine the beam within the same radius could also be used.)

## 2.4 Transverse Cooler

The IDS baseline cooling channel design consists of a sequence of identical 1.5 m long cells (figure 4). Each cell contains two sets of two 0.25 m-long rf cavities, with 1.5cm thick LiH absorbers blocks at the ends of each cavity set (4 per cell) and a 0.25 m spacing between cavities, and alternating solenoidal focusing coils. The LiH provides the energy loss material for ionization cooling. The total length of the cooling section is ~80m (50 cells plus matching). Based on simulations, the cooling channel reduces the transverse emittances by a factor of 2.5 in each dimension.

The cells contain two solenoidal coils with the coils containing opposite sign currents. The coils produce an approximately sinusoidal variation of the magnetic field on axis in the channel with a peak value on-axis of ~2.8T, providing transverse focusing with $\beta_\perp \cong 0.8$m.

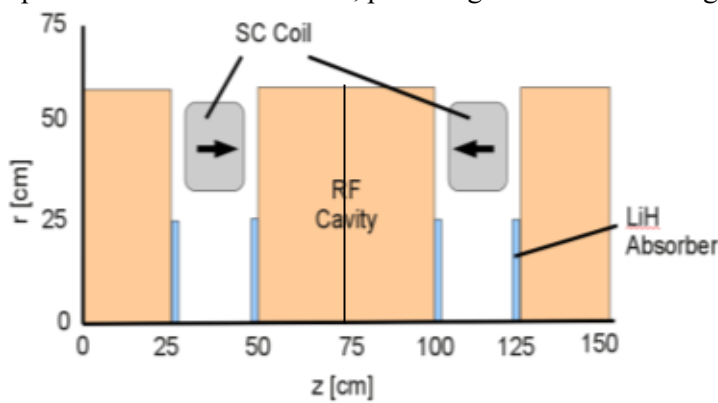

**Figure 4.** Layout of one period of the Alternating Solenoid cooling lattice, showing the alternating solenoid coils, the rf cavities and LiH absorbers.

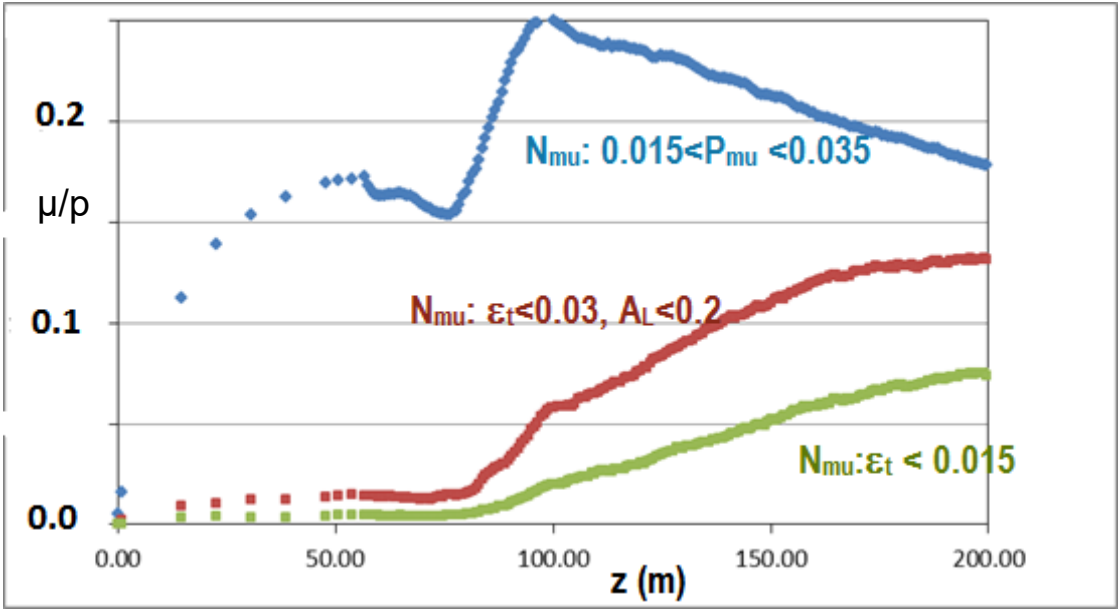

**Figure 5:** ICOOL Simulation results of the capture of muons within the front end system. All muons ($\mu^+$) within the momentum band are shown in blue. $\mu+$ (per 8 GeV p) within the acceptances are shown in red, half acceptance shown in green. The Rotator starts at z=75m and places beam within the momentum spread and acceptance cuts (z=100m). The cooler (z>100m) increases beam within the acceptance cuts while collimating much of the beam outside the acceptance.

ICOOL [14] simulation results of this front end system are shown in figure 5. At the end of the cooler (z= ~175m), simulations indicate that there are ~0.1 $\mu^+$ and $\mu^-$ per initial ~8 GeV proton within the projected acceptance of the downstream muon accelerator. (Particles with amplitudes $\varepsilon_{x,y}$ < ~0.03 m and longitudinal amplitudes $A_L$ <0.2 m are considered to be within that acceptance.) Simulations have also been performed using G4beamline [15] with similar results.

**2.5 Energy loss and activation**

The transport line from the target is designed to accept a maximal number of secondaries and therefore includes a large flux of particles that are not accepted into the final muon beams. These particles are mostly lost in the walls of the vacuum chamber (at up to 1 kW/m), potentially causing unacceptably large activation. This includes a large flux of protons of all energies (up to the primary proton beam energy), as well as many pions and electrons. To control and localize these losses a chicane and absorber system can be inserted into the transport downstream of the target [14].

The chicane consists of a double bent solenoid, with a central solenoidal field of 2.0T. The first solenoid bends outward ~21.7° over 6.5 m (bend radius 17.2m) The second 6.5m bend solenoid bends back by 21.7° to remain parallel. This displaces the beamline by ~0.92m. The net effect of the chicane is to remove higher momentum particles (P > ~500MeV/c) while transmitting particles within the final momentum acceptance with relatively small perturbations and losses. Following the chicane, and a further ~30m drift to accommodate $\pi \rightarrow \mu$ decay, the beam passes through an absorber (10cm thick Be, see figure 6). The absorber removes protons, pions, electrons, and lower energy muons.

The total effect of the chicane and absorber is to localize most losses and activation into the first ~50m of transport that includes the chicane and absorber. The downstream transport containing rf and cooling systems has greatly reduced activation losses (to < 10 W/m). With the rematching of the bunching and energy rotation, the total length of the Front End is increased by ~30m, with the difference largely obtained from the additional drift for π decay between chicane and absorber. The chicane/absorber system also reduces the number of muons within the final acceptance by ~10%. The parameters of the chicane and absorber have been chosen to reduce the downstream losses while maintaining the muon acceptance as much as possible; further optimization and improvement is possible.

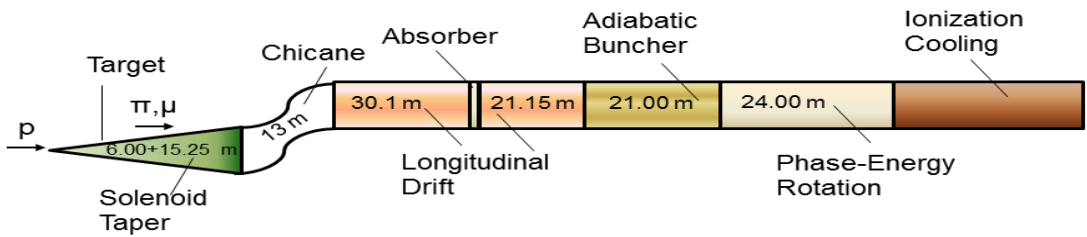

**Figure 6.** Layout of the Front End with a chicane and absorber.

## 2.6 rf breakdown considerations

The baseline design requires the use of high-gradient rf (~325MHz, 15MV/m) in moderately high magnetic fields (~2T). Initial experiments and analysis showed that rf gradients could be limited in high magnetic fields, and it was uncertain that the baseline parameters could be achieved [15, 16]. More recent results using carefully prepared rf cavity surfaces have shown greater tolerance for magnetic fields. 800 MHz rf cavities operated at 20 MV/m within 5T magnetic fields in recent experiments [17].

Experiments have also shown that gas-filled rf cavities suppress breakdown, even within high magnetic fields and the presence of rf accelerating beams [18]. Variations of the Front End using gas-filled rf cavities were developed and found to obtain capture and acceptance approximately equal to vacuum cavity examples, provided the rf gradients are increased to compensate energy loss in the gas [19]. Operation of gas-filled rf has a complication in that beam loading by electrons from beam ionization can drain rf power; this can be moderated by adding a small fraction of electronegative dopant ($O_2$) to the $H_2$, to facilitate electron recombination [20, 21].

With a gas-filled system and stronger rf gradients the Rotator can provide a significant amount of muon cooling, reducing the requirements for downstream cooling and providing a more compact muon source. Other cooling channel designs are also gas-filled [22, 23, 24, 25].

The original focusing fields were limited to 2T, partially out of concern for the rf gradient limitations. If those limitations are relaxed, a larger focusing field (3 - 4T) would enable a more efficient capture.

# 3. Cooling channel design

The simplified cooler design initially presented for the Front End has some significant deficiencies. It only cools transversely, and the beam is longitudinally heated throughout the channel, implying losses throughout the channel and also implying the downstream cooling and acceleration systems must have the same (or better) longitudinal acceptance. Alternative cooling systems are being considered; the currently preferred alternative cooling system is the "HFOFO Snake".

## 3.1 "HFOFO Snake" cooling channel

The "HFOFO Snake" was designed to obtain simultaneous transverse and longitudinal cooling, for both $\mu^+$ and $\mu^-$. The HFOFO snake is based on two principles: alternating solenoid focusing and resonant dispersion generation by a helical perturbation [7, 8]. In a homogeneous longitudinal magnetic field the two transverse modes are the cyclotron and drift modes [26]. Ionization losses cool only the cyclotron mode. Changing the solenoid polarity exchanges the identity of the cyclotron and drift modes, so that in an alternating solenoid lattice both modes are damped. (This principle is also used in the initial baseline 4-D cooling channel.) In the HFOFO channel the solenoids are tilted in a periodic helical pattern, generating a helical closed orbit perturbation with dispersion. This perturbation introduces a path length increase with momentum (positive momentum compaction) through the absorbers, which yields a longitudinal cooling effect. Wedges matched to the dispersion can give an additional cooling effect.

After some parameter variation and optimization, an HFOFO solution for the Front End was generated [27]. One period of the channel is shown in Fig. 7. It consists of:

- 6 Alternating solenoids (coil parameters: $L = 30$ cm, $R_{in} = 42$ cm, $R_{out} = 60$ cm) placed with a period of 70 cm along the axis. With current density 94.6 A/mm$^2$ the solenoids provide focusing with betatron phase advance ≈74°/step at a muon momentum of 230 MeV/c. To create a transverse magnetic field component, the solenoids are periodically inclined in rotating planes at $x \cdot \cos(\phi_k) + y \cdot \sin(\phi_k) = 0$, $\phi_k = \pi(1-2/N_s)(k+1)$, $k = 1, 2, \ldots, N_s$. $N_s$ is the number of solenoids/period. $N_s = 6$ in the present case. The rotation angles $\phi_k$ are: $4\pi/3$, 0, $2\pi/3$, $4\pi/3$, 0, $2\pi/3$; $\phi = 0$ corresponds to a tilt in the vertical plane. The chosen pitch angle of 2.5 mrad is too small to be visible in the figure.

- 6 Paired RF cavities ($f_{RF} = 325$ MHz, $L = 2 \times 25$ cm, $E_{max} = 25$ MV/m) filled with H$_2$ gas at a density that is 20% of liquid hydrogen, with Be windows. The radius and thickness of the Be windows are reduced in 3 steps along the channel: $R_w = 30$ cm, $w = 0.12$ mm (first 10 periods), $R_w = 25$ cm, $w = 0.10$ mm (next 10 periods) and $R_w = 20$ cm, $w = 0.07$ mm (last 10 periods). (Slab LiH absorbers could be used to replace the H$_2$, if vacuum-filled rf is preferred.)

- LiH wedge absorbers, providing additional longitudinal cooling. Although the momentum compaction factor of the lattice is positive, it is not sufficient for longitudinal cooling and wedges must be added. The wedge angle smoothly varies along the channel from 0.17 rad to 0.20 rad. The tip of the wedge just intercepts the channel axis so the muons on the equilibrium orbit (see figure 7) traverse no more than 0.3 mm of LiH.

An important feature of the design is that $\mu^-$ in solenoids 4, 5, 6 see exactly the same forces as $\mu^+$ in solenoids 1, 2, 3 and vice versa, so that $\mu^-$ and $\mu^+$ orbits have exactly the same form, with a longitudinal shift by a half period (three solenoids), and are not mirror-symmetric as one might expect. This allows us to find an orientation of the wedge absorbers (with periodicity = 2) such that they provide longitudinal cooling for both $\mu^-$ and $\mu^+$.

The complete cooling channel contains 30 periods for a total length of 126 m. Cost estimates for earlier cooling channels may be found in reference [28]. RF power supplies are a large fraction of the cost. Magnetrons might be more efficient and less expensive if phase and power control can be improved [29].

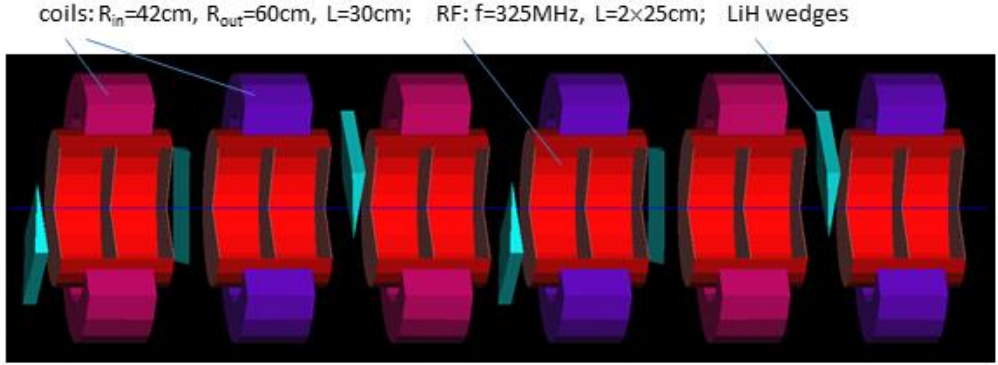

**Figure 7.** Layout of one period of the HFOFO lattice, showing the alternating solenoid coils (violet and blue), the rf cavities (red) and wedge LiH absorbers (cyan).

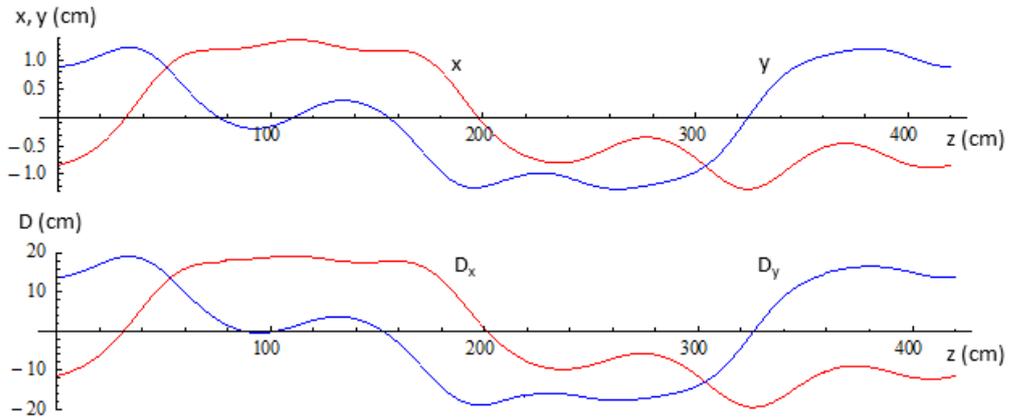

**Figure 8.** Closed orbit(above) and dispersion (below) in one period of the HFOFO lattice.

The closed orbit through the period plus the dispersion (for $\mu^+$) is shown in figure 8. The closed orbit approximately follows a circle of radius ~1.2cm, and the dispersion follows a ~20cm amplitude circle through the helical orbit. The normal mode tunes of the lattice are ($\nu_1$, $\nu_2$, $\nu_3$) = (1.227+0.010i, 1.2375+0.0036i, 0.1886+0.0049i), where the first two modes are transverse and the last is predominantly longitudinal; the imaginary part indicates the damping from ionization cooling.

The HFOFO channel was matched in closed orbit and betatron functions to the output of the Front End Rotator, using perturbations of the first 9 HFOFO solenoid strengths and tilts. In the HFOFO, the central momentum is gradually reduced from the ~250 MeV/c of the Rotator to ~210 MeV/c at the end of the HFOFO cooler.

Beam from the Front End Rotator was matched into the HFOFO cooler and tracked using the G4beamline simulation code. Results of the simulation are presented in figures 9 and 10 [27]. Both $\mu^+$ and $\mu^-$ are cooled. The rms transverse emittance $\varepsilon_t = (\varepsilon_1 \varepsilon_2)^{1/2}$ is reduced from ~16 mm to ~2.6mm, which is a factor of 6.2. Longitudinal emittance is reduced from ~24 mm to ~7.4 mm, a factor of 3.2. Total 6-D cooling is a factor of 123. Beam survival of the core muon beams (from decay and aperture losses) was ~70%.

The transverse cooling is greater than that of the previous 4-D cooling system. This is in part due to the stronger focussing and longer channel length. Initial losses are ~10% greater, partly from the more complicated 6-D phase space match, and partly from aperture losses in adding the helical orbit to the transverse motion. However the additional longitudinal cooling enables the use of the longer channel and provides a better match into downstream systems [22, 23, 30, 31], which are likely to have more limited longitudinal acceptances.

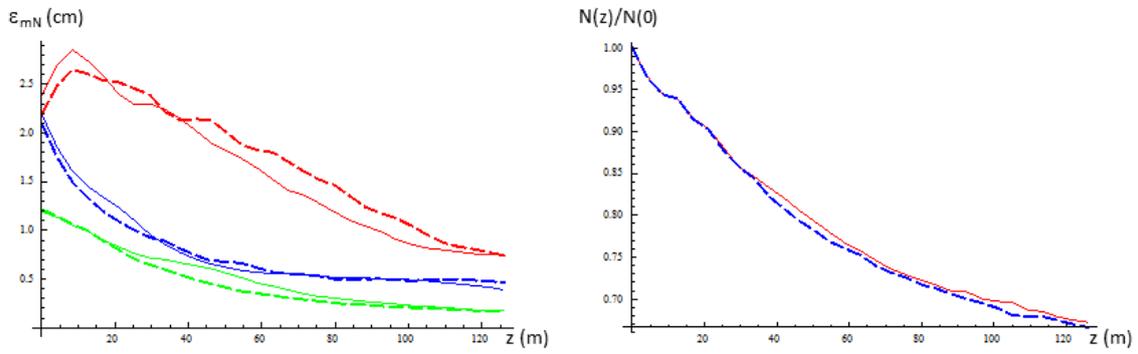

**Figure 9.** Results of beam simulations of cooling. (Left)Evolution of the eigenemittances ($\varepsilon_1$, $\varepsilon_2$, $\varepsilon_3$: green, blue, red) through the 124m cooling channel (solid lines are $\mu^+$, dashed are $\mu^-$, red). (Right) Beam survival within the transport apertures.

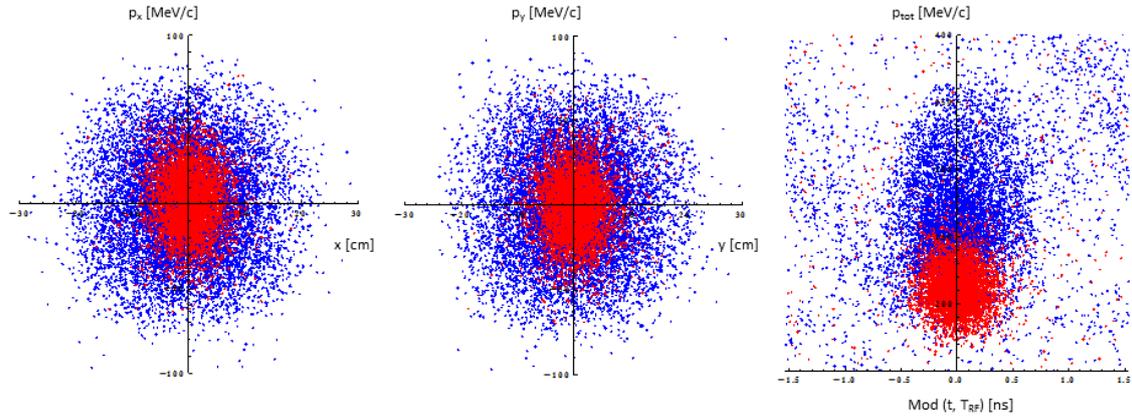

**Figure 10.** Results of beam simulations of cooling. Phase space distributions of the initial μ+ beam (blue) and of the cooled beam in the exit solenoid (red). (x-$p_x$, y-$p_y$, p-ct) All bunches were projected onto the same RF bucket in the rightmost plot. No cuts were applied.

## 4. Discussion and Summary

The results here present a baseline design for the muon Front End as developed by the MAP collaboration. It would provide a large-acceptance and capture system that could provide an intense source of bunched and cooled µ⁺ and µ⁻. The example presented is certainly not a final, optimized design and the research has suggested many possibilities for future improvement and modification.

The 2T baseline focussing is not optimum; stronger focusing of ~3T would provide better matching of beam into the rf and would readily match into a stronger focusing cooling channel. If the research confirms that the rf is compatible with the higher field, the Front End should be modified to match.

In the present paper we have displayed two cooling channels: an alternating solenoid channel that cools only transversely and a "HFOFO Snake" that cools both transversely and longitudinally. In the present development, the HFOFO channel is preferred because of the addition of longitudinal cooling.

The present channel was developed for muon collider/neutrino factory applications. It may be enough cooling for a neutrino factory. The present methods could also be adapted to obtain intense low-energy muon sources.

## Acknowledgments

We thank the MAP collaboration for support and assistance in this research. Fermilab is operated by Fermi Research Alliance, LLC under Contract No. DE-AC02-07CH11359 with the U. S. Department of Energy.